\documentclass[a4paper
,12pt]{article}
\usepackage{graphicx}
\begin{document}

\title{\bf Pattern Matching via Choice Existential Quantifications in Imperative Languages}
\author{Keehang Kwon \\
\sl \small Dept. of Computer Engineering, DongA  University\\
\sl \small 840 hadan saha, Busan, Korea\\
\small khkwon@dau.ac.kr
 }
\date{}
\maketitle

\newenvironment{describe}{\begin{list}{}{\setlength\leftmargin{80pt}}\setlength\labelsep{10pt}\setlength\labelwidth{70pt}}{\end{list}}

\newenvironment{flag}{\begin{list}{\makebox[20pt]{\hss$\circ$\enspace}}
                                  {\labelwidth 20pt}}{\end{list}}

%% js \newtheorem{proposition}{Proposition}

%% js \newenvironment{proof}
     %% js {\begin{trivlist}\item[]{\bf Proof. }}%
     %% js {\\* \hspace*{\fill} $\Box$\end{trivlist}}

\newenvironment{numberedlist}
{\begin{list}{\makebox[20pt]{\hss(\arabic{itemno})\enspace}}
             {\usecounter{itemno}\labelwidth 20pt}}{\end{list}}

\newenvironment{alphabetlist}
{\begin{list}{\makebox[20pt]{\hss(\alph{itemno1})\enspace}}
             {\usecounter{itemno1}\labelwidth 20pt}}{\end{list}}

\newenvironment{romanlist}
{\begin{list}{\makebox[20pt]{\hss(\roman{itemno2})\enspace}}
             {\usecounter{itemno2}\labelwidth 20pt}}{\end{list}}

\newcounter{itemno}

\newcounter{itemno1}

\newcounter{itemno2}
\newcounter{lemma}
\newcounter{exno}

\newcounter{defno}

%\newcounter{exno}[section]

%\newcounter{defno}[section]

%\newtheorem{defn}{Definition}[section]

%\newtheorem{ex}[defn]{Example}

%% js \newtheorem{lemma}{Lemma}

%% js \newtheorem{theorem}[lemma]{Theorem}

\newenvironment{defn}{\refstepcounter{defno}\medskip \noindent {\bf
Definition \thedefno.\ }}{\medskip}

\newenvironment{ex}{\refstepcounter{exno}\medskip \noindent {\bf
Example \theexno.\ }}{\medskip}

\newenvironment{millerexample}{
 \begingroup \begin{tabbing} \hspace{2em}\= \hspace{5em}\= \hspace{5em}\=
\hspace{5em}\= \kill}{
 \end{tabbing}\endgroup}

\newenvironment{wideexample}{
 \begingroup \begin{tabbing} \hspace{2em}\= \hspace{10em}\= \hspace{10em}\=
\hspace{10em}\= \kill}{
 \end{tabbing}\endgroup}

\newcommand{\sep}{\;\vert\;}

\newcommand{\ra}{\rightarrow}
\newcommand{\app}{\ }
\newcommand{\appt}{\ }
\newcommand{\tup}[1]{\langle\nobreak#1\nobreak\rangle}

\newcommand{\hu}{{\cal H}^+}
\newcommand{\Free}{{\cal F}}
\newcommand{\oprove}{\vdash\kern-.6em\lower.7ex\hbox{$\scriptstyle O$}\,}
\newcommand{\true}{\top}

\newcommand{\Dscr}{{\cal D}}
\newcommand{\Pscr}{{\cal P}}
\newcommand{\Gscr}{{\cal G}}
\newcommand{\Fscr}{{\cal F}}
\newcommand{\Vscr}{{\cal V}}
\newcommand{\Uscr}{{\cal U}}
\newcommand{\pderivation}{{\cal P}\kern -.1em\hbox{\rm -derivation}}
\newcommand{\pderivationl}{{\cal P}\kern -.1em\hbox{\em -derivation}}
\newcommand{\pderivable}{{\cal P}\kern -.1em\hbox{\rm -derivable}}
\newcommand{\pderivablel}{{\cal P}\kern -.1em\hbox{\em -derivable}}
\newcommand{\pderivations}{{\cal P}\kern -.1em\hbox{\rm -derivations}}
\newcommand{\pderivability}{{\cal P}\kern -.1em\hbox{\rm -derivability}}
\newcommand{\eqm}[1]{=_{\scriptscriptstyle #1}}
\newcommand\subsl{\preceq}
\newcommand{\fnrestr}{\uparrow}

\newcommand{\match}{{\rm MATCH}}
\newcommand{\triv}{{\rm TRIV}}
\newcommand{\imit}{{\rm IMIT}}
\newcommand{\proj}{{\rm PROJ}}
\newcommand{\simpl}{{\rm SIMPL}}
\newcommand{\failed}{{\bf F}}

\newcommand{\Dsiginst}[1]{{[#1]_\Sigma}}
\newcommand{\Psiginst}[1]{{[#1]_\Sigma}}
\newcommand{\lnorm}{{\lambda}norm}
\newcommand{\seq}[2]{#1 \supset #2}
\newcommand{\dseq}[2]{#1_1,\ldots,#1_{#2}}

\newcommand{\all}{\forall}
\newcommand{\some}{\exists}
\newcommand{\lambdax}[1]{\lambda #1\,}
\newcommand{\somex}[1]{\some#1\,}
\newcommand\allx[1]{\all#1\,}

\newcommand{\subs}[3]{[#1/#2]#3}
\newcommand{\rep}[3]{S^{#2}_{#1}{#3}}
\newcommand{\ie}{{\em i.e.}}
\newcommand{\eg}{{\em e.g.}}

% These are the annotations used with inference figures
\newcommand{\lbotr}{$\bot$-R}
\newcommand{\ldbotr}{\bot\mbox{\rm -R}}
\newcommand{\landl}{$\land$-L}
\newcommand{\ldandl}{\land\mbox{\rm -L}}
\newcommand{\landr}{$\land$-R}
\newcommand{\ldandr}{\land\mbox{\rm -R}}
\newcommand{\lorl}{$\lor$-L}
\newcommand{\ldorl}{\lor\mbox{\rm -L}}
\newcommand{\lorr}{$\lor$-R}
\newcommand{\ldorr}{\lor\mbox{\rm -R}}
\newcommand{\limpl}{$\supset$-L}
\newcommand{\ldimpl}{\supset\mbox{\rm -L}}
\newcommand{\limpr}{$\supset$-R}
\newcommand{\ldimpr}{\supset\mbox{\rm -R}}
\newcommand{\lnegl}{$\neg$-L}
\newcommand{\ldnegl}{\neg\mbox{\rm -L}}
\newcommand{\ldnegr}{\neg\mbox{\rm -R}}
\newcommand{\lalll}{$\forall$-L}
\newcommand{\ldalll}{\forall\mbox{\rm -L}}
\newcommand{\lallr}{$\forall$-R}
\newcommand{\ldallr}{\forall\mbox{\rm -R}}
\newcommand{\lsomel}{$\exists$-L}
\newcommand{\ldsomel}{\exists\mbox{\rm -L}}
\newcommand{\lsomer}{$\exists$-R}
\newcommand{\ldsomer}{\exists\mbox{\rm -R}}
\newcommand{\ldlamlr}{\lambda}
\newcommand{\sequent}[2]{\hbox{{$#1\ \longrightarrow\ #2$}}}
\newcommand{\prog}[2]{\hbox{{$#1\ \supset\ #2$}}}
\newcommand{\run}{\Gamma}

\newcommand{\Ibf}{{\bf I}}
\newcommand{\Cbf}{{\bf C}} 
\newcommand{\Cbfpr}{{\bf C'}}

\newcommand{\cprove}{\vdash_C}
\newcommand{\iprove}{\vdash_I}

\newsavebox{\lpartfig}
\newsavebox{\rpartfig}

% From the hohh section

\newenvironment{exmple}{
 \begingroup \begin{tabbing} \hspace{2em}\= \hspace{3em}\= \hspace{3em}\=
\hspace{3em}\= \hspace{3em}\= \hspace{3em}\= \kill}{
 \end{tabbing}\endgroup}
\newenvironment{example2}{
 \begingroup \begin{tabbing} \hspace{8em}\= \hspace{2em}\= \hspace{2em}\=
\hspace{10em}\= \hspace{2em}\= \hspace{2em}\= \hspace{2em}\= \kill}{
 \end{tabbing}\endgroup}

\newenvironment{example}{
\begingroup  \begin{tabbing} \hspace{2em}\= \hspace{3em}\= \hspace{3em}\=
\hspace{3em}\= \hspace{3em}\= \hspace{3em}\= \hspace{3em}\= \hspace{3em}\= 
\hspace{3em}\= \hspace{3em}\= \hspace{3em}\= \hspace{3em}\= \kill}{
 \end{tabbing} \endgroup }

\newcommand{\sand}{sand} % choice disjunction
\newcommand{\pand}{pand} % choice disjunction
\newcommand{\cor}{cor} % choice disjunction

\newcommand{\lb}{\langle}
\newcommand{\rb}{\rangle}
\newcommand{\pr}{prov}
\newcommand{\prG}{intp}
\newcommand{\prSG}{intp_E}
\newcommand{\intp}{intp_o}
\newcommand{\prove}{exec} % choice conjunction
\newcommand{\np}{invalid} % choice conjunction
\newcommand{\Ra}{\supset}  
\newcommand{\add}{\oplus} % choice disjunction
\newcommand{\adc}{\&} % choice conjunction
\newcommand{\Cscr}{{\cal C}}
\newcommand{\seqweb}{SProlog}
\newcommand{\sprog}{{SProlog}}

\newtheorem{theorem}[lemma]{Theorem}

\newtheorem{proposition}[lemma]{Proposition}

\newtheorem{corollary}[lemma]{Corollary}
\newenvironment{proof}
     {\begin{trivlist}\item[]{\it Proof. }}%
     {\\* \hspace*{\fill} \end{trivlist}}

\newcommand{\seqand}{\prec}
\newcommand{\seqor}{\cup}
\newcommand{\seqandq}[2]{\prec_{#1}^{#2}}
\newcommand{\parandq}[2]{\land_{#1}^{#2}}
\newcommand{\exq}[2]{\exists_{#1}^{#2}}
\newcommand{\ext}{intp_G}

\newcommand{\muprolog}{{Java$^{choo}$}}
\newcommand{\uch}{choose}
\newcommand{\kch}{choose}
\renewcommand{\pr}{ex}
\renewcommand{\prove}{ex} % choice conjunction
\newcommand{\iif}{if-then-else}
\newcommand{\swi}{switch}
\newcommand{\try}{try-catch}
%\begin{document}

\noindent {\bf Abstract}: 
Selection statements -- if-then-else, switch and
try-catch -- 
are commonly used  in modern imperative programming languages. 
We propose another selection   statement called a {\it choice existentially quantified statement}. 
This statement 
turns out to be quite useful for pattern matching among several merits.  
Examples will be provided for this  statement.
%\end{summary}

{\bf keywords:} selection, pattern matching, choice quantification, print.
%\end{keywords}

\section{Introduction}
Most imperative languages have selection statements to control execution flow.
A selection statement allows the machine to 
 choose  between two or more statements during execution.
Selection statements typically include \iif\  and \try.
 Unfortunately, these statements are $not$ sufficient for expressing nondeterministic tasks in a concise way.

To ovecome these problems,  inspired by the work in \cite{Jap03,Jap08}, we propose a new kind of
  selection  statements called {\it choice existentially quantified statements} (CEQ statements).
This statement is quite simple and of the form

\[ \kch(x) G \]
\noindent
where  $G$ is a statement. This has the following execution semantics:

\[ \pr(\Pscr, \kch(x) G)\ if\
 \pr(\Pscr,  [t/x] G) \] 

\noindent where the term (or the value) $t$ is  chosen  by the machine and $\Pscr$ is a 
set of procedure (and function) definitions.
In the above definition,  the machine chooses a $successful$ term $t$
 and then proceeds
with executing $[t/x]G$. 

We also introduce a variant of the above, $\kch(x\in S)\ G$, which is called
a {\it bounded choice existentially quantified statement} (BCEQ statement).
Bounded quantifiers differ from unbounded quantifiers in that bounded quantifiers restrict the range of 
the variable $x$ to the set $S$. Thus, bounded quantifiers make it easier for the machine
 to choose a successful term.

It can be easily seen that our new statement subsumes
the $print$ statement. For example, let $G$ be a statement and let $E$ be an expression.
Then $G; print(E)$ can be converted to

\[ \kch(x) (G; x == E) \]

\noindent provided $x$ does not appear free in $G$ and the choice of $x$ is  visible to the user.
In the above, note that a boolean condition is  a legal statement
in our language, as we shall see in Section 2.

The CEQ statement makes it simple to represent complex, nondeterministic tasks. For example, the following
statement represents the task of finding (and printing) an index $x$ (between 1 and 50) 
such that  the $x$th Fibonacci number is 5. 

\[ \kch(x\in \{ 1..50 \}) ( 5 == fib(x)) \]

\noindent In this case, the machine
will find the value of 6 for $x$ after some search.

Another example is the following. This statement  represents the task of finding and printing the values
of the tenth Fibonacci number and the factorial of 20.  

\[ \kch(x)\ \kch(y) ( x == fib(10);  y == fact(20)) \]

\noindent  Note that the  above program is  compact and
 easy to read.

This paper focuses on the 
core of Java. This is to present the idea  in a concise way.
The remainder of this paper is structured in the following way. We describe 
the core Java with the CEQ statements
 in Section 2. In Section \ref{sec:modules}, we
present some example of  \muprolog.
Section~\ref{sec:conc} concludes the paper.

\section{The Language}\label{sec:logic}

The language is a subset of the core (untyped) Java
 with some extensions. It is described
by $G$- and $D$-formulas given by the syntax rules below:
\begin{exmple}
\>$G ::=$ \>   $ A \sep cond \sep x = E \sep  G;G \sep   \kch(x) G \sep  \kch(x\in S) G $ \\   \\
\>$D ::=$ \>  $ A = G\ \sep \all x\ D$\\
\end{exmple}
\noindent
In the above, $cond$ represents a boolean condition, $S$ represents a set and $E$ is an expression.
$A$  represents a head of an atomic procedure definition of the form $p(x_1,\ldots,x_n)$ where
each $x_i$ is a variable.
A $D$-formula  is called a  procedure (and function) definition.

In the transition system below, $G$-formulas will act as the
main program (or statements), and a set of $D$-formulas enhanced with the
machine state (a set of variable-value bindings) will act as  a program.

 We will  present an execution
semantics  via a proof theory \cite{Khan87,MNPS91,HM94,MN12}.
The rules  defines what
it means to
execute the main task $G$ from a program $\Pscr$.
These rules define precisely what is a success and failure. 
Below the notation $D;\Pscr$ denotes
$\{ D \} \cup \Pscr$ but with the $D$ formula being distinguished
(marked for backchaining). Note that execution  alternates between
two phases: the main phase (the phase of executing the main program)
and the backchaining phase (one with a distinguished clause).
The notation $S\ sand\ R$ denotes the following: execute $S$ and execute
$R$ sequentially. It is considered a success if both executions succeed.

\begin{defn}\label{def:semantics}
Let $G$ be a main task and let $\Pscr$ be a program.
Then the notion of   executing $\lb \Pscr,G\rb$ successfully and producing a new
program $\Pscr'$-- $ex(\Pscr,G,\Pscr')$ --
 is defined as follows:
\begin{numberedlist}

\item    $ex((A = G_1);\Pscr,A)$ if
 $ex(\Pscr, G_1)$ and  $ex(D;\Pscr, A)$.

\item    $ex(\all x D;\Pscr,A)$ if   $ex([s/x]D;
\Pscr, A)$ where $s$ is a value (or a term). \% argument passing

\item    $ex(\Pscr,A)$ if   $D \in \Pscr$ and $ex(D;\Pscr, A)$. \% a procedure call

\item  $ex(\Pscr,cond,\Pscr)$ if $eval(\Pscr,cond,cond')$ and $cond'$ is true. \% evaluating boolean 
condition $cond$ to $cond'$.

%\item  $ex(\Pscr,\neg cond,\Pscr)$ if $cond$ is false. \% boolean condition

\item  $ex(\Pscr,x=E,\Pscr\uplus \{ \lb x,E' \rb \})$ if $eval(\Pscr,E,E')$.
\% the assignment statement. If evaluating $E$ fails, then
the whole statement fails. Here, 
$\uplus$ denotes a set union but $\lb x,V\rb$ in $\Pscr$ will be replaced by $\lb x,E' \rb$.

\item  $ex(\Pscr,G_1; G_2,\Pscr_2)$  if $ex(\Pscr,G_1,\Pscr_1)$  sand 
  $ex(\Pscr_1,G_2,\Pscr_2)$.

\item $ex(\Pscr, \kch (x) G, \Pscr_1)$  if 
$ex(\Pscr, [t/x]G,\Pscr_1)$ where $t$ is a successful value for $x$  
 chosen by the machine.

\item $ex(\Pscr, \kch (x\in S) G, \Pscr_1)$ if $x \in S$ and 
$ex(\Pscr, [t/x]G,\Pscr_1)$ where $t$ is a successful value for $x$  
 chosen by the machine.

\end{numberedlist}
\end{defn}

\noindent
If $ex(\Pscr,G,\Pscr_1)$ has no derivation, then the machine returns  the failure.

\section{Examples }\label{sec:modules}

Pattern matching is a useful feature  in modern programming languages.
While there have been several attempts to add pattern matching to imperative paradigm \cite{RPS10},
these attempts are rather complex and rely on refining the type system.
The simplest approach to adding pattern matching, which requires no type systems, 
is to  allow  first-order terms as data. For example,
$tuple(tom,31,male)$ would be a legimate data.
In such a case, our $\kch$ statement is well-suited for
pattern matching. For example, 
the following statement is a simple implementation of
destructuring an employee's record into three components. 

\begin{exmple}
        $getrecord(emp)$ \{ \\
\> $\kch(name) \kch(age) \kch(sex)\  (tuple(name, age, sex) == emp)$; 
\end{exmple}

\noindent It is not easy to write concise codes for this task in
traditional languages.
Fortunately, it is quite simple in our setting.

\section{Conclusion}\label{sec:conc}

In this paper, we have considered an extension to a core Java with a new selection statement.
 This extension allows   $\kch (x) G$  where $x$ is a variable and $G$ is a statement.
This statement makes it possible for the core Java
to perform nondeterministic tasks.  
Our language gives, in a sense, a logical status to Java. This means that other  logical connectives such as disjunctions can be
 added. Some progress has been made towards this direction \cite{KHP13}.

\section{Acknowledgements}

This work  was supported by Dong-A University Research Fund.

\bibliographystyle{ieicetr}

\begin{thebibliography}{1}

\bibitem{Khan87}
G.~Kahn,  ``Natural Semantics'', In the 4th Annual Symposium on Theoretical Aspects of Computer Science, 
LNCS vol. 247,  1987.



\bibitem{Jap03}
G.~Japaridze, ``Introduction to computability logic'', Annals  of Pure and
 Applied  Logic, vol.123, pp.1--99, 2003.

\bibitem{Jap08}
G.~Japaridze,   ``Sequential operators in computability logic'',
 Information and Computation, vol.206, No.12, pp.1443-1475, 2008.  

\bibitem{KHP13}
K.~Kwon, S.~Hur and M.~Park,  ``Improving Robustness via Disjunctive Statements in Imperative  Programming'', IEICE Transations on Information and Systems, vol.E96-D,No.9, September, 2013.  

\bibitem{HM94}
J.~Hodas and D.~Miller,   ``Logic Programming in a Fragment of Intuitionistic Linear Logic'', 
 Information and Computation, vol.110, No.2, pp.327-365, 1994. 



\bibitem{MNPS91}
D.~Miller, G.~Nadathur, F.~Pfenning, and A.~Scedrov, ``Uniform proofs as a
  foundation for logic programming'', Annals of Pure and Applied Logic, vol.51,
  pp.125--157, 1991.

\bibitem{MN12}
D.~Miller, G.~Nadathur, Programming with higher-order logic, Cambridge University Press,   2012.


\bibitem{RPS10}
S. Ryu, C. Park and G. Steel Jr.  ``Adding Pattern Matching to Existing Object-Oriented
Languages'', FOOL '10, Nevada, USA,  2010.
\end{thebibliography}

%\profile*{}{}% without picture of author's face

%\end{multicols}

\end{document}